\documentclass{elsart}
\usepackage{epsfig}

\newcommand{\be}{\begin{equation}}
\newcommand{\ee}{\end{equation}}
\newcommand{\bc}{\begin{center}}
\newcommand{\ec}{\end{center}}
\newcommand{\bi}{\begin{itemize}}
\newcommand{\ei}{\end{itemize}}
\newcommand{\ba}{\begin{eqnarray}}
\newcommand{\ea}{\end{eqnarray}}

\begin{document}

\begin{frontmatter}

\title{Boundary effects in extended dynamical systems}

\author{V\'{\i}ctor M. Egu\'{\i}luz,
Emilio Hern\'andez-Garc\'{\i}a and Oreste Piro}
\address{Instituto Mediterr\'aneo de Estudios Avanzados, IMEDEA (CSIC-UIB)\\
E-07071 Palma de Mallorca, Spain}
\begin{abstract}
In the framework of spatially extended dynamical systems, we
present three examples in which the presence of walls lead to
dynamic behavior qualitatively different from the one obtained in
an infinite domain or under periodic boundary conditions. For a
nonlinear reaction-diffusion model we obtain boundary-induced
spatially chaotic configurations. Nontrivial average patterns
arising from boundaries are shown to appear in spatiotemporally
chaotic states of the Kuramoto-Sivashinsky model. Finally, walls
organize novel states in simulations of the complex
Ginzburg-Landau equation.
\end{abstract}
\date{7 November 1999}
\end{frontmatter} 
\smallskip


\section{Introduction.}

Most of our current knowledge of spatiotemporal chaos comes from
its analysis in the infinite size limit or from simulations in
finite domains with periodic boundary conditions\cite{CrossHoh}.
However, different geometries or boundary conditions may lead to
substantially different dynamical behavior. We will exemplify this
assertion by showing results from three different extended
dynamical systems in which the dynamics is strongly influenced by
the boundaries.

\begin{figure}[t]
  \epsfig{figure=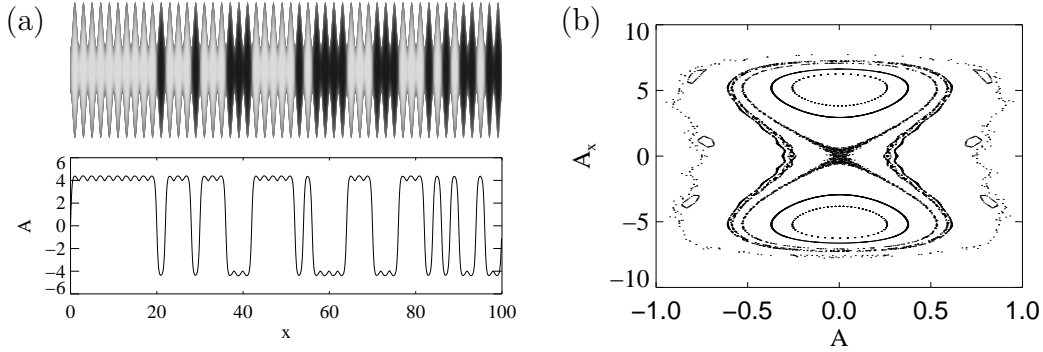, width=1.\textwidth}
  \put(-190,130){(b)}
  \put(-400,130){(a)}
  \caption{
  {\small (a) Top: two-dimensional configuration of our reaction-diffusion
  equation starting
  from random initial conditions. Bottom: Amplitude at the
  center of the domain.
  (b) Stroboscopic Poincar\'e map of the phase space of an
  approximation to our model equation.
  }}
\end{figure}

\section{Frozen spatial chaos induced by boundaries}
\label{frozen}

We first show how the presence of nontrivial boundaries can induce
the appearance of {\sl spatial chaos} in a system for which no
chaotic behavior is found neither in the infinite size limit nor
with purely periodic boundary conditions\cite{Eguiluz99b}. The
model we consider is a nonlinear diffusion equation of the
Fisher-Kolmogorov type: $\partial_t A = \nabla^2 A +A-A^3$. The
real quantity $A=A(x,y,t)$ is a twodimensionally extended field.
When solved in doubly periodic integration domains, regions in
which $A\approx
\pm 1$ form, grow, and compete until one of the two phases
takes over the whole system. When solved in regions such that
Dirichlet (that is $A=0$) conditions are applied in lateral
boundaries which are not straight but undulating (see Fig. 1) the
result is different: Frozen states in which the $A=+1$ and $A=-1$
phases alternate in space become stable and attract most of the
initial conditions. The alternation of the two phases is random
and produces static but spatially chaotic configurations. The
justification of the 'chaotic' adjective can be done with
different dynamical systems tools. For example Fig. 1b shows a
Poincar\'{e} map of some of the spatial configurations obtained
from an approximation to our model equation. KAM tori and other
fractal structures are evident, in direct analogy with the
classical picture of Hamiltonian systems with chaotic time
trajectories. Theoretical arguments can be developed to show that
the effect of the spatially undulated boundaries on the spatial
pattern is similar to time-periodic parametric forcing in common
temporal dynamical systems, from which the above chaotic
phenomenology can be understood. Further details are given in
\cite{Eguiluz99b}

\begin{figure}[t]
  \epsfig{figure=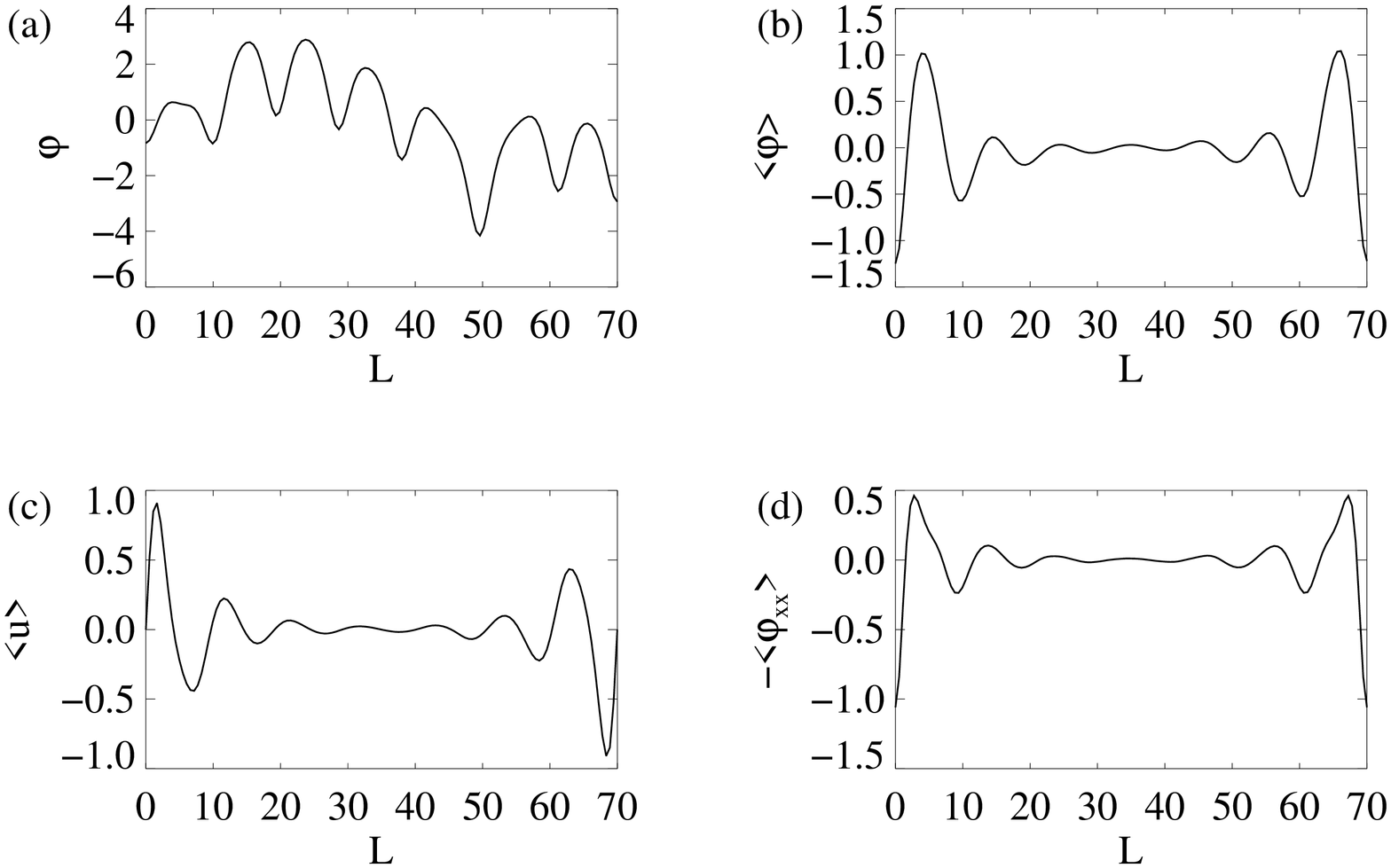, height=1.\textwidth,angle=90,clip=true}
  \caption{
  {\small   (a) A
  characteristic configuration of the one-dimensional
  Kuramoto-Sivashinsky equation with stress-free boundary conditions.
  (b) The time averaged field.
  }}
\end{figure}
\section{Average patterns of spatiotemporal chaos.}
\label{averages}

Chaotic pattern dynamics in many experimental systems
\cite{Gluckman95,Ahlers} show structured
time averages. In this second Section, we suggest that simple
universal boundary effects underlie this phenomenon and exemplify
them with the Kuramoto-Sivashinsky equation in a finite domain.
Figure~2b shows a structured average pattern for the
Kuramoto-Sivashinsky equation in one dimension
($\dot\varphi=-\partial^2_x
\varphi-\partial^4_x \varphi + (\partial_x \varphi)^2$) with
stress-free boundary conditions (null first and third spatial
derivatives at the boundaries). In contrast, the strong
fluctuations of the instantaneous field are also shown in Fig.~2a
for comparison. As in the experiments, the average pattern
recovers the symmetries which are respected by both the equation
and the boundary conditions (in this case left-right symmetry)
locally broken in the chaotic field. The amplitude is strongest at
the boundaries and decays through the center of the average
pattern. The strength of the oscillations in the average pattern
follows a $L^{-1/2}$ dependence on system size. Plateaus in the
average-pattern wavenumber as a function of the system size are
observed\cite{Eguiluz99}.  Most of these observations are also
found in experimental systems\cite{Gluckman95,Ahlers} for which
the Kuramoto-Sivashinsky equation does not apply, thus indicating
its generic, mainly geometrical, origin: what is relevant for
these phenomena to appear is the occurrence of strong enough
chaotic fluctuations in the presence of non-trivial boundaries.

\begin{figure}[t]
\begin{center}
  \epsfig{figure=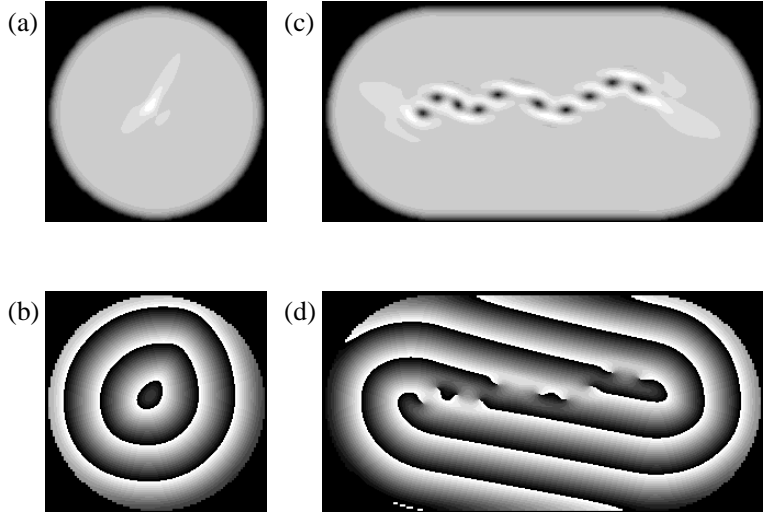, width=.8\textwidth,angle=0}
  \caption{{\small
   Frozen structures for the complex Ginzburg-Landau equation
  in the Benjamin-Feir stable regime under null Dirichlet boundary conditions in
  different domains. Top raw: modulus; bottom raw: phase
  }}
\end{center}
\end{figure}
\section{The complex Ginzburg-Landau equation in bounded domains.}
\label{scgle}

In our third example, the effect of a finite geometry on the
two-dimensional complex Ginzburg-Landau equation (in the
Benjamin-Feir stable regime\cite{CrossHoh}) is addressed. Boundary
conditions induce the formation of novel states. For example
target like-solutions (Fig.~3a-b) appear as robust solutions under
Dirichlet boundary conditions, whereas they are not observed under
periodic boudary conditions. Dirichlet boundary conditions play a
double r\^ole as sources (or sinks) of defects and as emitters of
plane waves. The interplay between these two properties of the
boundaries gives rise to interesting
behavior\cite{Eguiluz99a,Eguiluz99c}. In a square, walls emit
waves that develop shock lines when they cross. Spiral defects
form chains anchored by these shock lines. In circular domains,
however, the emission is definitively dominated by the internal
spiral defects. In a stadium geometry (Figs.~3c-d), typically a
chain of defects links the centers of the circular regions.
Synchronization of boundary wave emission is also
found\cite{Eguiluz99a,Eguiluz99c}. Most of these phenomena can be
understood from the emission properties of Dirichlet
walls\cite{Eguiluz99c}.

\section{Conclusions.}

We have shown three examples where the boundary conditions play a
key r\^ole in extended dynamical systems. Further work is needed
to clarify the degree of universality of these results and to find
general properties of different kinds of boundaries.

Financial support from DGICyT projects PB94-1167 and
PB97-0141-C02-01 is greatly acknowledged.


\newpage


\end{document}